\documentclass[namedreferences]{kluwer}
\usepackage[]{graphicx}

\begin{document}
\begin{article}
\begin{opening}
\title{Precision Spectroscopy \\at Heavy Ion Ring Accelerator SIS300}
\author{Hartmut \surname{Backe}}
\runningauthor{Hartmut Backe} \runningtitle{Precision
Spectroscopy} \institute{Institut f\"ur Kernphysik, Johannes
Gutenberg-Universit\"{a}t Mainz, D-55099 Mainz, Germany}
\date{January 03, 2007}

\begin{abstract}
Unique spectroscopic possibilities open up if a laser beam
interacts with relativistic lithium-like ions stored in the heavy
ion ring accelerator SIS300 at the future Facility for Antiproton
and Ion Research FAIR in Darmstadt, Germany. At a relativistic
factor $\gamma$ = 36 the $^{2}P_{1/2}$ level can be excited from
the ${}^{2}S_{1/2}$ ground state for any element with frequency
doubled dye-lasers in collinear geometry. Precise transition
energy measurements can be performed if the fluorescence photons,
boosted in forward direction into the X-ray region, are
energetically analyzed with a single crystal monochromator. The
hyperfine structure can be investigated at the
$^{2}P_{1/2}-{}^{2}S_{1/2}$ transition for all elements and at the
$^{2}P_{3/2}-{}^{2}S_{1/2}$ transition for elements with $Z\leq
50$. Isotope shifts and nuclear moments can be measured with
unprecedented precision, in principle even for only a few stored
radioactive species with known nuclear spin. A superior relative
line width in the order of $5\cdot10^{-7}$ may be feasible after
laser cooling, and even polarized external beams may be prepared
by optical pumping.
\end{abstract}
\keywords{laser spectroscopy, relativistic lithium-like ions,
laser cooling, hyperfine spectroscopy, nuclear polarization,
SIS300}

\abbreviations{\abbrev{FAIR}{Facility for Antiproton and Ion
Research}; \abbrev{HI}{Heavy Ion}; \abbrev{QED}{Quantum
Electrodynamics}; \abbrev{CCD}{Charge Coupled Device};
\abbrev{EBIT}{Electron Beam Ion Trap}}

\end{opening}


\section{Introduction} \label{introduction}
The central part of the planned FAIR project in Darmstadt is a
heavy ion synchrotron called SIS300 with a magnetic rigidity
$B\rho$ = 300 Tm and a circumference of 1100 m
[\citeauthor{Gutbrod06} \citeyear{Gutbrod06}]\footnote{The
parameters used in this paper do not match exactly with the
parameters of this report.}. The magnetic field will be produced
by superconducting magnets with a maximum induction of 6 Tesla
which can be ramped with a rate of 1 T/s. With this synchrotron
bare uranium can be accelerated up to a maximum energy of 34
GeV/u, corresponding to a relativistic factor
$\gamma=1/\sqrt{1-\beta^{2}} = 37.5$ and a reduced velocity
$\beta=v/c=0.9996444$, with $c$ the speed of light. A fascinating
possibility of this accelerator is the excitation of few electron
systems by the interaction with the light of conventional lasers
in a collinear geometry. If the laser beam counter-propagates
lithium-like uranium with $\gamma = 36$, the laser light at the
blue edge of the visible spectral range with an energy of
$\hbar\omega_{L}$ = 3.898 eV is Doppler-shifted and appears in the
rest frame of the Li-like system with an energy of
$\hbar\omega_{0} \cong 2 \gamma \hbar\omega_{L}$ = 280.6 eV. As
will be outlined in more detail in section \ref{lithium} this is
just the $^{2}P_{1/2}-{}^{2}S_{1/2}$ transition energy in
lithium-like uranium. A variety of spectroscopic possibilities
exist if this transition can be induced. As will be pointed out in
section \ref{precision} the combination with a precise single
crystal X-ray spectrometer, which detects the fluorescence photons
boosted in forward direction up to an energy of $\hbar\omega_{X}=2
\gamma \hbar\omega_{0}$ = 20.2 keV allows both, very accurate
measurements of the transition energy $\hbar\omega_{0}$ and the
relativistic factor $\gamma$. If radioactive Li-like ions could be
injected into the SIS300, a hyperfine spectroscopy would be
possible for radioactive species with nuclear spin $I>0$, see
section \ref{hyperfine}. A few remarks on laser cooling will be
made in section \ref{cooling}. In section \ref{pumping} the
possibility of a nuclear polarization by optical pumping with
circularly polarized laser light will briefly be touched on. The
paper closes in section \ref{conclusions} with a conclusion.

The essential ideas of this paper were for the first time
presented by the author of this paper in the year 2000 at GSI in
Darmstadt [\citeauthor{Backe00} \citeyear{Backe00}], see also
\citeauthor{Gutbrod01} \citeyear{Gutbrod01}.


\section{The $^{2}P_{1/2,3/2}-{}^{2}S_{1/2}$ transitions in lithium-like uranium}
\label{lithium}

\begin{figure}[t,b,h]
\centering
\includegraphics[scale=0.5,clip]{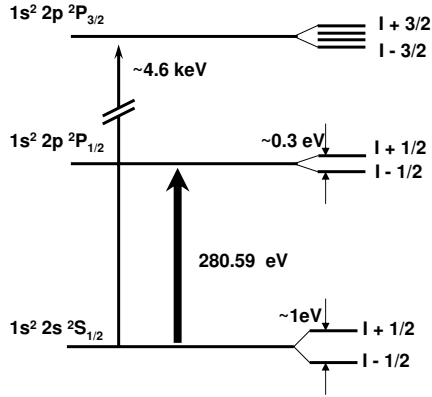}
\caption{The lithium-like level scheme of uranium} \label{level}
\end{figure}

The three electron, lithium-like level scheme of uranium is shown
in figure \ref{level}. The third electron outside the closed
$1s^{2}$ shell is a $2s$ electron, consequently the ground state
is a ${}^{2}S_{1/2}$ term. The lowest excited states belong to the
$1s^{2}2p$ configuration and form $^{2}P_{1/2}$ and $^{2}P_{3/2}$
terms. The large fine-structure splitting of about 4.3 keV
originates from relativistic effects.

A first precision measurement of the $^{2}P_{1/2} - {}^{2}S_{1/2}$
transition energy of (280.59 $\pm$ 0.09) eV was reported by
\citeauthor{Schweppe91} \citeyear{Schweppe91}. The aim of this
experiment was to test QED in few electron uranium. In a number of
publications \citeauthor{Lindgren94}, and \citeauthor{Persson93}
calculated (280.52$\pm$0.28) eV which was in good agreement with
the experiment. In the meantime better experiments by
\citeauthor{Brandau03}, who measured (280.516 $\pm$ 0.099) eV, and
by \citeauthor{Beiersdorfer05} were performed. In the latter
reference a value of $(280.645 \pm 0.015)$ eV is reported which
was obtained with SuperEBIT. The accuracy of previous experiments
was improved by nearly one order of magnitude. The best
calculation of \citeauthor{Yerokhin01} \citeyear{Yerokhin01}
without second order QED effects is (280.48 $\pm$ 0.11) eV, and
with inclusion of an estimated value of these effects (280.64
$\pm$ 0.21) eV [\citeauthor{Yerokhin05} \citeyear{Yerokhin05}],
i.e. the precision of the measurement exceeds currently the
precision of the calculation by a factor of 14. Despite this
situation a scheme is presented in the next section with which in
future the experimental precision probably can be further improved
by another factor of 4. Alternatively, it may be useful to measure
$\beta$ and $\gamma$ of the Li-like ions with accuracies in the
order of $10^{-8}$ and $10^{-5}$, respectively, see subsection
\ref{SubSectPrecision}.
\section{Precision transition energy measurement \\in Li-like uranium at SIS300}
\label{precision}
\subsection{Proposed experimental setup}
\begin{figure}[t,b]
\centering
\includegraphics[scale=0.5,clip]{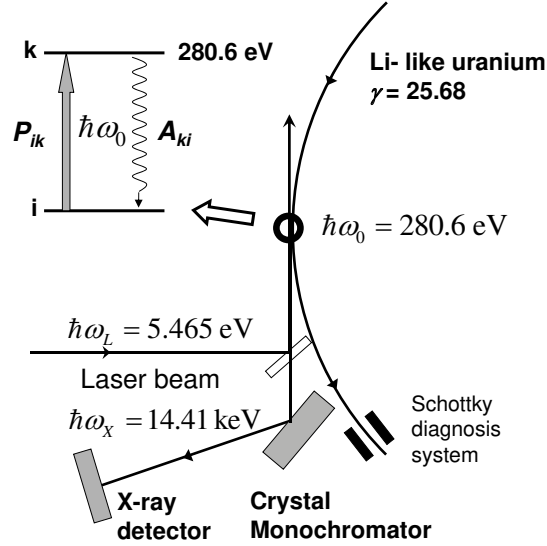}
\caption{Proposed schematic experimental setup. The laser beam
with photons of energy $\hbar\omega_L$ is reflected by a thin
mirror into a straight section of SIS300 and interacts in a
collinear geometry with lithium-like uranium ions moving in
opposite direction. De-excitation photons from the $^{2}P_{1/2} -
{}^{2}S_{1/2}$ transition with an energy $\hbar\omega_0 = 280.6$
eV appear predominantly in forward direction with respect to the
HI beam at an X-ray energy $\hbar\omega_X = 14.41$ keV. The photon
energy is measured by a single crystal monochromator.}
\label{setup}
\end{figure}

The proposed experimental setup is schematically shown in figure
\ref{setup}. A laser beam with a photon energy of 5.465 eV,
corresponding to a vacuum wavelength $\lambda$ = 226.87 nm,
counter-propagates in a straight section of SIS300 a Li-like
uranium beam with a relativistic factor $\gamma = 25.68$. The
photon energy in the rest frame of the Li-like U$^{89+}$-ion is
\begin{equation}\label{transition-energy}
\hbar\omega_{0} = \sqrt{\frac{1+\beta}{1-\beta}}~\hbar\omega_{L} =
280.6~\mbox{eV}
\end{equation}
which is with $\beta = \sqrt{1-1/\gamma^2} = 0.99924152$ just the
$^{2}P_{1/2} - {}^{2}S_{1/2}$ transition energy. De-excitation
photons emitted in forward direction with respect to the Li-like
uranium beam are boosted in the laboratory system to an energy
\begin{equation}\label{X-ray-energy}
 \hbar\omega_{X} = \sqrt{\frac{1+\beta}{1-\beta}}~\hbar\omega_{0} = 14.41~\mbox{keV}.
\end{equation}
Occurrence of X-rays indicates resonance absorption of laser
photons in the Li-like system. However, for an accurate
measurement of the transition energy $\hbar\omega_0$ also the
reduced velocity $\beta$ or the relativistic factor $\gamma$ must
be known with high precision. This can be achieved by an energy
measurement of the X-ray photons with the aid of a single crystal
monochromator. In this manner, two very precise experimental
methods are combined. The laser photon energy can be determined
easily with a relative accuracy of $5\cdot10^{-7}$ and the X-ray
energy with $2.8\cdot10^{-5}$, see subsection
\ref{SubSectPrecision}. Combining equations
(\ref{transition-energy}) and (\ref{X-ray-energy}), the transition
energy $\hbar\omega_{0}$, the reduced velocity $\beta$, and the
the relativistic factor $\gamma$ of the Li-like uranium ion can be
determined:
\begin{eqnarray}
 \hbar\omega_{0} &=& \sqrt{\hbar\omega_{X}\cdot\hbar\omega_{L}}~,\label{energy}\\
 \beta &=&\frac{\hbar\omega_X/\hbar\omega_L - 1}{\hbar\omega_X/\hbar\omega_L +
 1}~,\label{beta}\\
 \gamma &=& \frac{1}{2}\frac{\hbar\omega_X/\hbar\omega_L +
 1}{\sqrt{\hbar\omega_X/\hbar\omega_L}}
 \cong\frac{1}{2}\sqrt{\frac{\hbar\omega_{X}}{\hbar\omega_{L}}}\label{gamma}.
\end{eqnarray}
The relative accuracies of a measurement of $\beta$ and $\gamma$
are $\Delta\beta/\beta = 2.1\cdot 10^{-8}$ and
$\Delta\gamma/\gamma = 1.4\cdot 10^{-5}$, respectively, see also
subsection \ref{SubSectPrecision}.\footnote{It should be mentioned
that the precision of the X-ray energy measurement and also the
current value of the $^{2}P_{1/2} - {}^{2}S_{1/2}$ transition
energy $\hbar\omega_0$ is not high enough for an improved test of
the time dilatation in special relativity. Such a test is based on
equation (\ref{energy}) which can be rewritten with a small
additional term as
$\hbar\omega_L\cdot\hbar\omega_X/\hbar\omega_0^2 = 1 + 2
\hat{\alpha}(\beta^{2}+...)$. The upper limit of the parameter
$\hat{\alpha}$ is currently $\hat{\alpha}<2.2\cdot 10^{-7}$, see
\citeauthor{Saathoff03} \citeyear{Saathoff03}. An improvement of
this value would require a measurement of $\hbar\omega_L$,
$\hbar\omega_X$, and $\hbar\omega_0$ with at least an accuracy of
$10^{-7}$.}
\subsection{Angular distribution and photon energy in the laboratory system}\label{SubSectAngDist}
In reality the photons are emitted in the rest frame of the
Li-like ion with an angular distribution which will be assumed to
be isotropic, i.e. $d\dot{n}_{0}/d\Omega_{0} =\dot{n}_{0}/(4\pi)$.
The angular distribution in the laboratory frame is given by
\begin{eqnarray}
 \frac{d\dot{n}}{d\Omega} &=&
 \frac{d\dot{n}_{0}}{d\Omega_{0}}\frac{d\Omega_{0}}{d\Omega}\frac{dt_{0}}{dt}~,\label{angular-distribution}\\
 \frac{d\Omega_{0}}{d\Omega} &=&
 \frac{1}{\gamma^{2}(1-\beta\cdot\cos\Theta)^{2}}~,\label{dOmegaTransform}\\
 \frac{dt_{0}}{dt} &=& \frac{1}{\gamma}\label{dtTransform}
\end{eqnarray}
with $d\Omega_{0}/d\Omega$ the relativistically transformed solid
angle ratio which follows from the relation
\begin{equation}\label{angleTransform}
 \cos\Theta_0 = \frac{\cos\Theta-\beta}{1-\beta\cos\Theta}~.
\end{equation}
Here $\Theta_0$ and $\Theta$ are the observation angles with
respect to the velocity vector $\textbf{v}$ of an individual
Li-like ion in the rest frame of the Li-like ion and the
laboratory frame, respectively. Further on, $dt_{0}/dt = 1/\gamma$
is the relativistic time dilatation\footnote{For a nice survey of
relevant formulas of Lorentz transformation in storage rings see
\citeauthor{Habs91} \citeyear{Habs91}, and \citeauthor{Schramm04}
\citeyear{Schramm04}}. The transition energy in the rest frame of
the Li-like system is
\begin{equation}\label{hbaromega0}
 \hbar\omega_{0} = \gamma(1+\beta\cos\Psi)\hbar\omega_{L}
\end{equation}
with $\Psi$ the angle which an individual ion with the velocity
vector $\textbf{v}$ makes with the laser beam axis. In the small
angle approximation, with $\overrightarrow{\theta}$ the
observation angle with respect to the nominal velocity vector
$\textbf{v}_0$ of the Li-like ion beam and $\overrightarrow{\psi}$
the angular deviation of an individual ion from $\textbf{v}_0$,
the photon energy is
\begin{equation}\label{hbaromegaX}
 \hbar\omega_{X} = \frac{\hbar\omega_{0}}
 {\gamma(1-\beta\cdot\cos|\overrightarrow{\theta}-\overrightarrow{\psi}|)}
 \cong\frac{2\gamma}{1+(|\overrightarrow{\theta}-\overrightarrow{\psi}|\gamma)^2}\hbar\omega_{0}~,
\end{equation}
with $\hbar\omega_0$ the photon energy in the Li-like system. From
equations (\ref{hbaromega0}) and (\ref{hbaromegaX}) the relation
\begin{equation}\label{energy-distribution}
 \hbar\omega_{X} = \frac{1+\beta\cos\psi}
 {1-\beta\cdot\cos|\overrightarrow{\theta}-\overrightarrow{\psi}|}\hbar\omega_{L}
 \cong\frac{4\gamma^2}{1+(|\overrightarrow{\theta}-\overrightarrow{\psi}|\gamma)^2}\hbar\omega_{L}~.
\end{equation}
follows which directly relates the laser photon energy
$\hbar\omega_{L}$ to the X-ray energy $\hbar\omega_{X}$.

\begin{figure}[t,b]
\centering
\includegraphics[scale=0.455,clip]{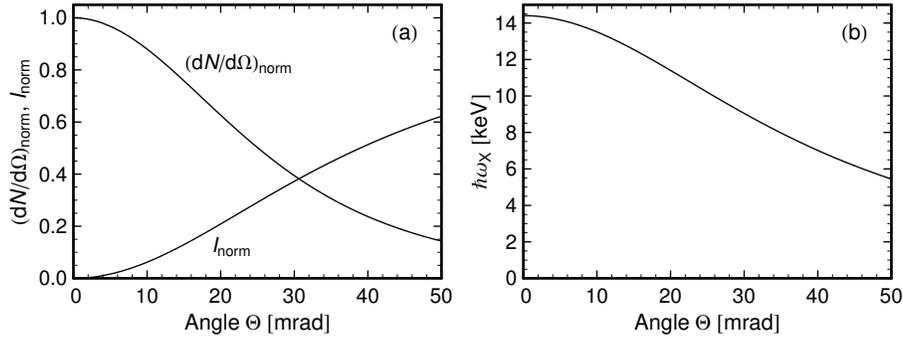}
\caption{(a) Angular distribution $(dN/d\Omega)_{norm}$ and
integrated intensity $I_{norm}=\int_0^{\Theta}
(dN/d\Omega)d\Omega/(4\pi)$, and (b) X-ray energy
$\hbar\omega_{X}$, both as function of the observation angle
$\Theta$ in the laboratory system for $\Psi$ = 0. The angular
distribution $(dN/d\Omega)_{norm}$ is normalized to its value at
$\Theta$ = 0, $I_{norm}$ approaches 1 for $\Theta\rightarrow\pi$.}
\label{ang-energy-distrib}
\end{figure}

Angular distribution $dN/d\Omega$, integrated intensity
$\int_0^{\Theta} (dN/d\Omega)d\Omega$ and X-ray energy
$\hbar\omega_{X}$ are shown in fig. \ref{ang-energy-distrib} (a)
and (b) as function of the observation angle $\Theta$. It is
worthwhile to notice that in a cone with a polar angle $\Theta =
50$ mrad, corresponding to 2.86$^{\circ}$ only, already more than
60 \% of the intensity is concentrated.
\subsection{The laser system and fluorescence rate estimate}
Laser light with a wavelength of $\lambda$ = 226.9 nm can be
produced by an Excimer laser running on XeF (351/353 nm) which
pumps a dye laser, for example. The output of the latter must be
frequency doubled by a BBO crystal. Assuming a mean output power
of the Excimer laser $\overline{P}$ = 250 W at a repetition rate
$f_{rep}$ = 10 kHz and a pulse width of $\Delta t_{pulse}$ = 10
ns, the instantaneous pulse power output of the frequency doubled
laser may amount to 50 kW. A bandwidth of the dye laser radiation
$\Delta\nu_{D} \simeq$ 1 GHz can be reached by means of an
intracavity etalon which may further be reduced in the frequency
doubling unit to $\Delta\nu_{L} \simeq$ 0.7 GHz or
$\Delta\nu_{L}/\nu_{L} \simeq 5.3\cdot 10^{-7}$. With these
numbers the spectral photon flux within a single laser pulse
amounts to $\Delta\dot{N}_{pulse}/(\Delta\nu_{L}/\nu_{L}) \simeq
1.1 \cdot 10^{29}$/s.

A lifetime of 61.8 $\pm$ 1.2(stat) $\pm$ 1.3(syst) ps of the 2p
$^{2}P_{1/2}$ has been measured by \citeauthor{Schweppe91} which
corresponds to a transition rate $A_{ki} = 1.62\cdot 10^{10}$/s or
a relative level width $\Gamma/\hbar\omega_{0} = 0.38\cdot
10^{-7}$. This width is small in comparison to the relative band
width of the photon flux in the rest frame of the Li-like ion
$\Delta\hbar\omega_{0}/\hbar\omega_{0} = \Delta\nu_{L}/\nu_{L}=
5.3\cdot 10^{-7}$. Under these circumstances the induced
transition rate is given by the equation
\begin{equation}\label{Pik}
 P_{ik} = \frac{d^2\dot{N}_{pulse}^{0}}{dA\cdot d
 \hbar\omega_{0}/\hbar\omega_{0}} A_{ki}
 \frac{g_k}{g_i}\frac{\pi^2}{c} \Big(\frac{\hbar c}{\hbar\omega_0}
 \Big)^3~,
\end{equation}
with $g_k$ and $g_i$ the statistical weights. At a cross-section
$A$ = 10 mm$^2$ of the laser beam in the interaction region, the
photon flux in the rest frame of the Li-like ion is
$d^2\dot{N}_{pulse}^{0}/(dA\cdot d\hbar\omega_{0}/\hbar\omega_{0})
= \gamma \cdot d^2\dot{N}_{pulse}/(dA\cdot d\nu_{L}/\nu_{L}) =
\gamma \cdot 1.1 \cdot 10^{28}$/(mm$^2$s). The number of induced
transitions in a laser pulse of duration $\Delta t_{pulse}$ = 10
ns is, with $P_{ik}$ of equation (\ref{Pik}), $P_{ik} \Delta
t^{0}_{pulse} = P_{ik} \Delta t_{pulse}/\gamma$ = 20.4, with
$\Delta t^{0}_{pulse}$ the pulse length in the rest frame of the
Li-like system. This number is much larger as the spontaneous
transition rate $A_{ki} \Delta t_{pulse}/\gamma$ = 6.3, meaning
that the effect of saturation must be taken into account. In the
following all estimations of the spontaneous transition rates are
performed in the saturation limit, since the experimental
conditions are close to saturation. The number of emitted photons
per laser pulse and per lithium-like uranium ion is then
\begin{equation}\label{Nsat}
 N_{0,sat}=\frac{1}{2}\Big(A_{ki}\frac{\Delta t_{pulse}}{\gamma}+1\Big) = 3.7~.
\end{equation}
The additional summand 1 in the brackets originates from the fact
that at saturation the Li-like ion is left with 50\% probability
in the excited state after the laser pulse has been past.
\subsection{The single crystal monochromator}
\begin{figure}[t,b]
\centering
\includegraphics[scale=0.5,clip]{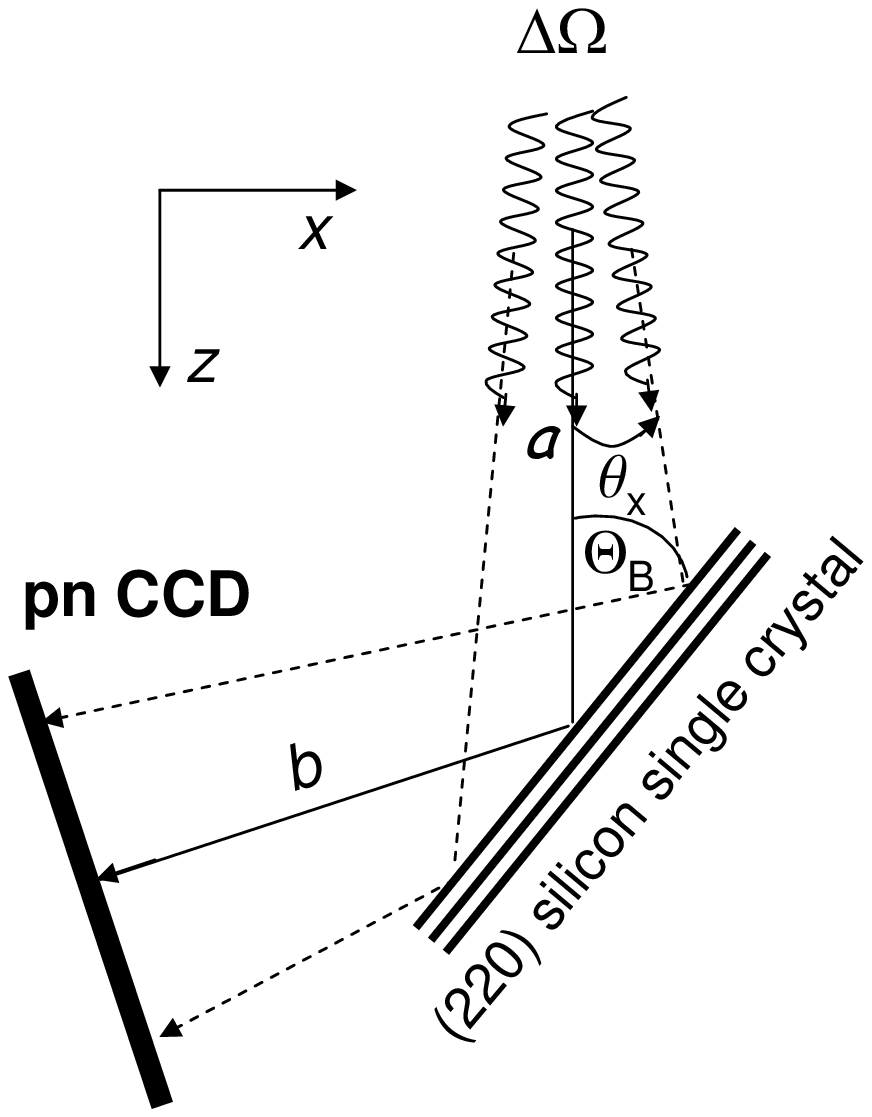}
\includegraphics[scale=0.45,clip]{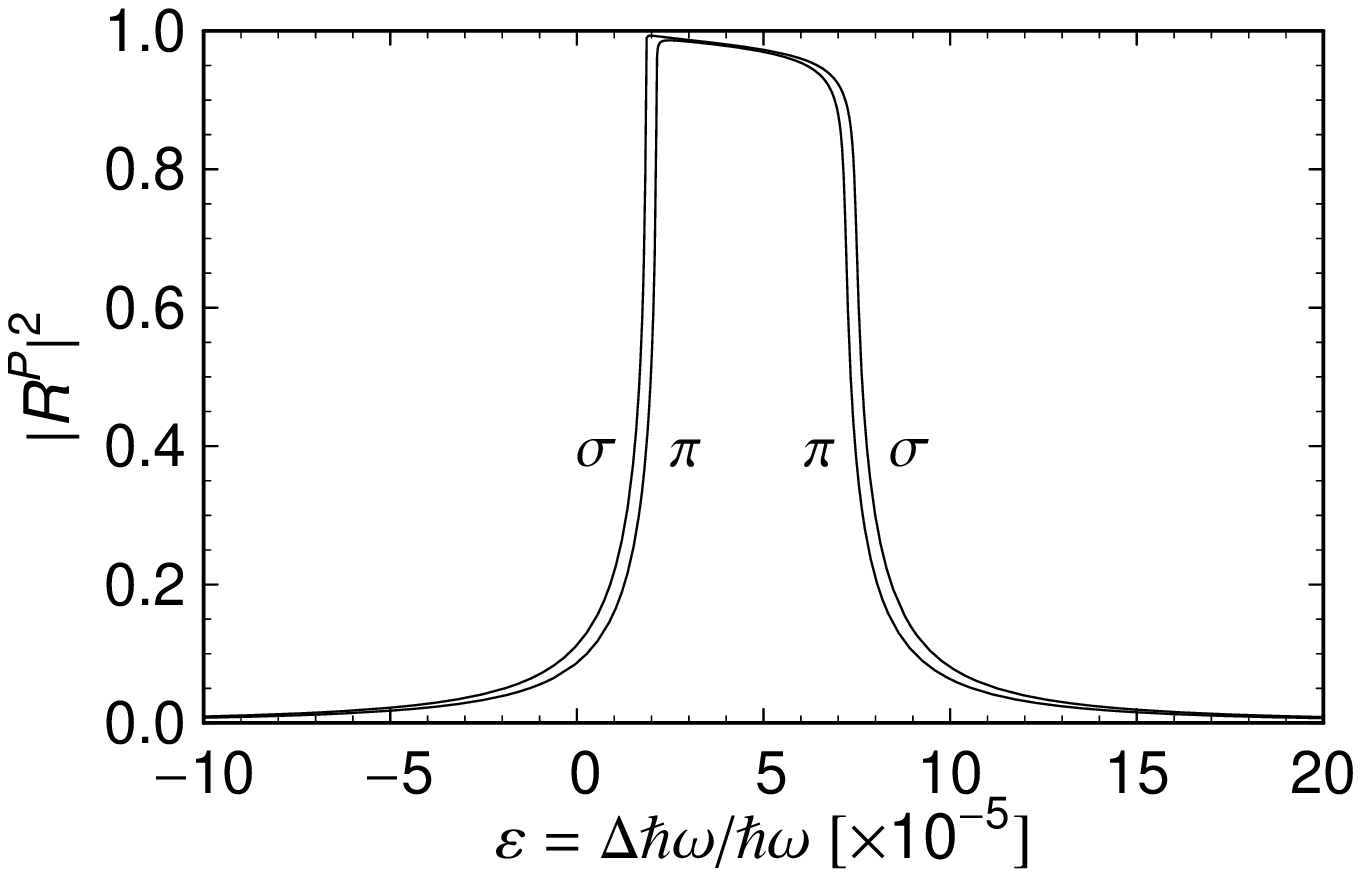}
\caption{Scheme of the monochromator (left), and reflecting power
ratios for $\pi$ and $\sigma$ polarized X-rays at $\theta_x$ = 0
(right). Dielectric susceptibilities $\chi^{\prime}_0 = -0.46776
10^{-5}$, $\chi^{\prime\prime}_0 = -0.34777 10^{-7}$,
$\chi^{\prime}_H = -0.28191 10^{-5}$, $\chi^{\prime\prime}_H =
-0.33364 10^{-7}$ were used in the calculation. A pn-CCD detector
serves as a position sensitive and energy dispersive detector.}
\label{monochromator}
\end{figure}

The monochromator is shown in figure \ref{monochromator}. A bent
silicon single crystal with its surface cut parallel to the (220)
crystal planes acts as a cylindrical mirror for the X-rays. It is
energy dispersive in the horizontal direction. The deviation
$\varepsilon$ of the photon energy from the nominal Bragg energy,
defined by the equation $\hbar\omega_{X} =
\hbar\omega_{B}(1+\varepsilon)$, with
\begin{eqnarray}
 \hbar \omega_{B}=\frac{2\pi \sqrt{h^2+k^2+l^2}}{a_{0}}\frac{\hbar c}{2\sin\Theta_{B}}~, \label{hw0}
\end{eqnarray}
is approximately given by the expression [\citeauthor{Caticha89}
\citeyear{Caticha89}]
\begin{eqnarray}
\varepsilon=\frac{\chi_0^{\prime}}{2\sin^2\Theta_B}-\frac{\theta_x}{\tan\Theta_B}~.\label{epsilon}
\end{eqnarray}
Here $\theta_x = \Theta-\Theta_B$ is the horizontal deviation from
the nominal Bragg angle $\Theta_B$. The integers $h,k,l$ are the
Miller indices, $a_{0}=5.4309~\rm \AA$ the lattice constant, and
$\chi^{\prime}_0$ the real part of the mean dielectric
susceptibility $\chi_0 = \chi^{\prime}_0 +
i\chi^{\prime\prime}_0$. The Bragg angle for $\hbar \omega_{B}$ =
14.413 keV amounts for the (220) reflection to $\Theta_{B}=
12.944^{\circ}$.

The finite energy width of the Bragg reflection can be calculated
from the reflecting power ratio ${|R^P|}^2$ with the amplitude
ratio given by [\citeauthor{Caticha89} \citeyear{Caticha89},
Eq.~(3.2)]
\begin{eqnarray}
R^P(\theta_x,\varepsilon)&=&-y_P(u)+\mbox{sign}[\Re{(y_P(u))}]\sqrt{y^2_P(u)-1} \label{rap} \\
y_P(u)&=&\frac{u+i\chi_0^{\prime\prime}}{P\chi_H} \label{ypu} \\
u &=& 2 \sin^2
\Theta_B\Big[\Big(1-\frac{a}{R\cdot\sin\Theta_{B}}\Big)\frac{\theta_x}{\tan\Theta_B}+\varepsilon\Big]
+\chi_0^{\prime}~. \label{u}
\end{eqnarray}
The quantity $\chi_H = \chi^{\prime}_H + i\chi^{\prime\prime}_H$
is the Fourier component of the dielectric susceptibility of the
analyzer crystal for the reciprocal lattice vector $\textbf{H}$,
$R$ the bending radius of the crystal, and $P$ the polarization
factor. The latter is $P=\cos2\theta_{B}$ for $\pi$ polarization,
with the polarization vector in the reflection plane, and $P=1$
for $\sigma$ polarization, with the polarization vector
perpendicular to the reflection plane. Corresponding reflecting
power ratios are shown in figure \ref{monochromator}. Parameters
for $\chi_0$ and $\chi_H$ were taken from \citeauthor{Stepanov97}
\citeyear{Stepanov97}.

The energy width of the Bragg reflection is in a good
approximation given by the solution of equation (\ref{u}) for
$\theta_{x}$ = 0 with $u-\chi^{\prime}_0 = \pm |P||\chi_{H}|$,
i.e. the relative energy width is $\Delta\varepsilon =
\Delta\hbar\omega_X/\hbar\omega_X = |P||\chi_{h}|/\sin^2\Theta_B$.
With the energy deviation $\varepsilon = -(\theta\gamma)^2$ of the
emitted X-rays as function of the emission angle
$\overrightarrow{\theta} = \textbf{e}_x \theta_x + \textbf{e}_y
\theta_y$, which follows from equation (\ref{energy-distribution})
for $\psi = 0$, one obtains from equation (\ref{u}) a quadratic
equation
\begin{equation}\label{quad-eqn}
 \theta_x^2+\theta_y^2\pm\Big(1-\frac{a}{R\cdot\sin\Theta_{B}}
 \Big)\frac{\theta_x}{\gamma^2\tan\Theta_B}
 -\frac{|P||\chi_{h}|}{2\gamma^2\sin^2\Theta_B} = 0
 \end{equation}
the solution of which describes the accepted angular region. For a
bending radius $R$ chosen such that $1-a/(R\sin\Theta_B) = 0$ is
fulfilled, the accepted angles $\theta_x$ and $\theta_y$ are
located within a circle of radius
$\sqrt{|P||\chi_{H}|/(2\gamma^2\sin^2\Theta_B)}$ and the accepted
solid angle is just $\Delta\Omega^P  =\pi|P||\chi_{H}|/
(2\gamma^2\sin^2\Theta_B)$. The corresponding solid angle in the
rest frame of the Li-like system is $\Delta\Omega_0^P = 4 \gamma^2
\Delta\Omega^P$. The sum of both polarization states, normalized
to $4\pi$, is
\begin{equation}\label{delta-omega0}
 \frac{\Delta\Omega_0}{4\pi} =
 \frac{(1+|\cos2\Theta_B|)|\chi_{H}|}{4\sin^2\Theta_B}~.
\end{equation}
With $|\chi_{H}| = 0.282\cdot10^{-5}$ [\citeauthor{Stepanov97}
\citeyear{Stepanov97}] the result is $\Delta\Omega_0/4\pi =
2.67\cdot10^{-5}$. The expected count rate at the pn-CCD detector
is at saturation with $\dot{n_{0}} = A_{ki}/2$
\begin{equation}\label{count-rate}
 \dot{n} =
 \frac{1}{2}\frac{A_{ki}}{\gamma}\frac{(1+|\cos2\Theta_B|)|\chi_{h}|}{4\sin^2\Theta_B}~.
\end{equation}
It should be mentioned that the focal length of the cylindrical
monochromator crystal amounts to $f = (R/2)\sin\Theta_B$ which
reduces with $1-a/(R\sin\Theta_B) = 0$ to $f = a/2$. It follows
from the image equation $1/a+1/b = 1/f$ that at $a = b$ the focus
at the pn-CCD detector is just a vertical line.
\subsection{Count rate estimate}
The detected X-ray rate at the pn-CCD detector is given by the
equation
\begin{equation}\label{pnCCD-count-rate}
 \dot{N}_X = \frac{\Delta\nu_L/\nu_L}{\Delta\gamma_{b}/\gamma}
 f_{rep}\cdot N_{Li}\cdot N_{0,sat}\frac{\Delta\Omega_0}{4\pi}\varepsilon_X~.
\end{equation}
The first factor is the fraction
$(\Delta\gamma_L/\gamma)/(\Delta\gamma_{b}/\gamma)$ of Li-like
ions in a bunch which can be pumped. With $\Delta\gamma_L/\gamma =
\Delta\nu_L/\nu_L$ which follows from equation (\ref{hbaromega0})
for $\Delta\hbar\omega_0 = 0$, it is given by the overlap of the
relative laser bandwidth $\Delta\nu_L/\nu_L = 5.3\cdot 10^{-7}$
with the relative energy spread $\Delta\gamma_{b}/\gamma= 10^{-4}$
of the Li-like ions in a bunch. A reduction due to the angular
spread $\sigma^{\prime}_{HI}$ of the Li-like beam can be neglected
as long as $\sigma^{\prime}_{HI} \ll 2 \sqrt{\Delta\nu_L/\nu_L} =
1.46$ mrad holds. This latter relation can also be derived from
the Doppler-shift formula (\ref{hbaromega0}) for which a first
order expansion in $\hbar\omega_L$ and a second order expansion in
$\Psi$ results in $\Delta\Psi =
2\sqrt{\Delta\hbar\omega_L/\hbar\omega_L}=
2\sqrt{\Delta\nu_L/\nu_L}$ for $\Delta\hbar\omega_0 = 0$. In
addition are $f_{rep} = 10^4$/s the laser repetition rate, which
must be synchronized with a circulating bunch, $N_{Li} = 10^5$ the
number of Li-like ions in a bunch, $N_{0,sat} = 3.7$, and
$\Delta\Omega_0/4\pi = 2.67\cdot10^{-5}$. The overall efficiency
$\varepsilon_X = 0.2$ takes into account the photon detection
efficiency of the pn CCD as well as photon absorption in the
window of the SIS300 vacuum chamber and the mirror for the laser
light. The result for the count rate according to equation
(\ref{pnCCD-count-rate}) is $\dot{N_X} = 104$/s which looks quite
reasonable.
\subsection{Precision of energy measurement} \label{SubSectPrecision}
The relative accuracy of a measurement of the
$^{2}P_{1/2}-{}^{2}S_{1/2}$ transition energy is, according to
equation (\ref{energy}), given by
\begin{equation}\label{accuracy}
 \frac{\Delta\hbar\omega_0}{\hbar\omega_0} =
 \frac{1}{2}\sqrt{\Big(\frac{\Delta\hbar\omega_{L}}{\hbar\omega_{L}}\Big)^2
 +\Big(\frac{\Delta\hbar\omega_{X}}{\hbar\omega_{X}}\Big)^2}.
\end{equation}
The precision of the X-ray energy measurement has two
contributions. One is a sort of statistical error which is assumed
to be 30 \% of the half width of the Bragg reflex, i.e.,
$\delta\hbar\omega_X/\hbar\omega_X \simeq 0.3
|\chi_{H}|/\sin^2\Theta_B = 1.7\cdot 10^{-5}$, with
$|\chi_{H}|/\sin^2\Theta_B = 5.6\cdot 10^{-5}$. The other one is a
systematical error which originates from the energy calibration.
Let us assume that the calibration is performed with the
14.41302(32) keV line of a $^{57}$Co ($t_{1/2}$ = 271 d) source
which for this purpose must be placed temporarily in the
interaction region of laser and Li-like ion beam. The relative
precision $\delta\hbar\omega_{14.4}/\hbar\omega_{14.4} =
2.2\cdot10^{-5}$ is of the same order of magnitude as the X-ray
energy measurement. Since the error of the laser frequency
measurement can be neglected, the total expected relative error of
the $^{2}P_{1/2}-{}^{2}S_{1/2}$ transition energy is
$\Delta\hbar\omega_0/\hbar\omega_0 = 1.4\cdot10^{-5}$ or
$\Delta\hbar\omega_0 = 0.0039$ eV. The latter would be a factor of
about 4 better as the above quoted value of
\citeauthor{Beiersdorfer05} \citeyear{Beiersdorfer05}.

According to equation (\ref{gamma}) the relativistic factor
$\gamma$ can be measured simultaneously with the same relative
precision $\Delta\gamma/\gamma = 1.4\cdot10^{-5}$. It should be
mentioned that the relative accuracy of $\beta$ is
$\Delta\beta/\beta = 2.1\cdot 10^{-8}$ because of
$\Delta\beta/\beta = (\Delta\gamma/\gamma)/(\beta\gamma)^2$ which
follows from $\gamma=1/\sqrt{1-\beta^{2}}$.
\section{Hyperfine spectroscopy} \label{hyperfine}
If the nuclear spin $I$ is not zero, ground and excited states of
Li-like ions exhibit a hyperfine splitting. The ${}^{2}S_{1/2}$
and the $^{2}P_{1/2}$ states are split only by the magnetic
hyperfine interaction while for the $^{2}P_{3/2}$ also the
quadrupole interaction contributes. The hyperfine splitting of the
${}^{2}S_{1/2}$ ground-state is typically in the order of 0.5 eV
[see, e.g., \citeauthor{Shabaev98} \citeyear{Shabaev98}, or
\citeauthor{Boucard00} \citeyear{Boucard00}]. It will be shown in
the following that a hyperfine structure of the
$^{2}P_{1/2}-{}^{2}S_{1/2}$ or even the
$^{2}P_{3/2}-{}^{2}S_{1/2}$ transition can be investigated by
laser spectroscopy as well.
\begin{figure}[t,b]
\centering
\includegraphics[scale=0.5,clip]{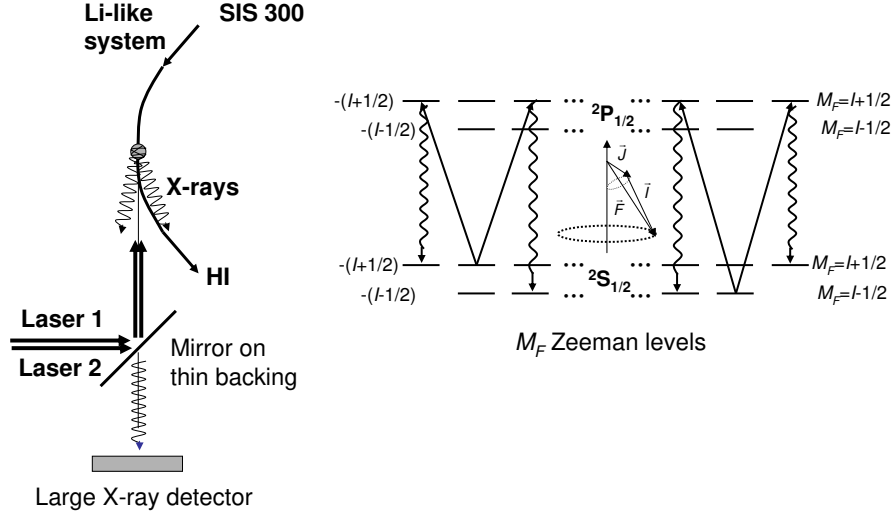}
\caption{Scheme of the experimental setup for a measurement of the
hyperfine structure (left), and pumping scheme (right). Shown are
the Zeeman levels of the ground and excited state with $F =
I\pm1/2$ the total angular momentum quantum numbers, and $I$ the
nuclear spin. While induced laser transitions can populate at
resonance only one excited hyperfine level, radiative (downward)
transitions can populate both ground state hyperfine levels. For
clarity, not all possible transitions are drawn in.}
\label{SetupHyperfine}
\end{figure}

The experimental setup is depicted in figure \ref{SetupHyperfine}.
It might be advantageous to use two laser beams, one pumping,
e.g., the $F_g = I-1/2\rightarrow F_e = I+1/2$ transition, and the
other the $F_g = I+1/2\rightarrow F_e = I+1/2$ one. Otherwise the
signal may cease rapidly because of depopulation pumping. The
relative line width $\Delta\hbar\omega_0/\hbar\omega_0 =
\Delta\gamma_{b}/\gamma$ of a transition is entirely determined by
the energy distribution of the Li-like ions in a bunch which is in
the order of $\Delta\gamma_{b}/\gamma = 10^{-4}$. Therefore, a
line width $\Delta\hbar\omega_0 \simeq 0.03$ eV is expected which
may be small enough to resolve the hyperfine pattern.

Since in such an experiment only small relative energy changes of
the hyperfine components must be measured, a high resolution X-ray
detector is not required. A large area detector placed in forward
direction can be used. This has the advantage that the accepted
solid angle can be increased by a large factor, see figure
\ref{ang-energy-distrib}. For an accepted polar angle of $\Theta$
= 4 mrad only, the relative accepted solid angle in the rest frame
of the Li-like system is $\Delta\Omega_0/(4\pi)$ = 0.0104, i.e.,
in the order of 1 \%. Consequently, the count rate is expected to
be rather high. Indeed, with
$(\Delta\nu_L/\nu_L)/(\Delta\gamma_{b}/\gamma) = 5.3\cdot10^{-3}$,
$f_{rep} = 10^4$/s, $N_{Li} = 1$, $N_{0,sat} = 3.7$,
$\Delta\Omega_0/(4\pi)$ = 0.0104, and $\varepsilon_X = 0.5$ the
count rate is according to equation (\ref{pnCCD-count-rate})
$\dot{N}_X = 1.0$/s. Notice, that in principle only one stored ion
is required for the envisaged hyperfine spectroscopy! However, it
is not at all clear that the X-ray detector may be located in
forward direction because of expected excess background count
rates. But even if a deflection of the X-ray beam out of the
forward direction is necessary, e.g., by a pyrolytic graphite
crystal which accepts a much larger bandwidth as a single crystal
monochromator, the count rate may be sufficiently large with a few
tens of Li-like ions in a bunch. Alternatively, the frequency of
the pulse laser may be increased. Thereby the event rate increases
linearly until the circulation frequency of the ions in the ring
of about 275 kHz has been reached. It might also be sufficient to
pump only with one laser beam and allow for depopulation pumping
if the count rate is high enough.

This kind of hyperfine spectroscopy at Li-like ions has a number
of advantages. First of all, hyperfine fields can be calculated
with a very high accuracy, at least with a much better accuracy as
for neutral atoms which is typically 10 \% for the isotope shift
and 3 \% for hyperfine fields. This fact is of great importance
since relative measurements can be avoided allowing a direct
access to the Bohr-Weisskopf effect. Secondly, isotope shifts and
magnetic hyperfine splittings can be measured for any element via
the $^{2}P_{1/2}-{}^{2}S_{1/2}$ transition. At the highest
relativistic factor $\gamma$ = 36 even a quadrupole interaction
can be studied via the $^{2}P_{3/2}-{}^{2}S_{1/2}$ transition for
all elements with $Z\leq50$. (For the $2s~{}^2S_{1/2}-2p~
^2P_{1/2,3/2}$ transition energies see \citeauthor{Bosselmann99}
\citeyear{Bosselmann99}, or \citeauthor{Johnson96}
\citeyear{Johnson96}.) Finally, it should be mentioned once more
that the sensitivity is very high and only a very few or even only
one radioactive ion may already be sufficient for the
spectroscopy. However, such experiments would require a
re-injection of radioactive species, produced by fragmentation
reactions, into SIS300 as Li-like ions.
\section{Laser cooling} \label{cooling}
The line width of the hyperfine components can be improved by
several orders of magnitude by laser cooling as will be
demonstrated in the following. \footnote{The subject of laser
cooling for relativistic beams has been discussed by
\citeauthor{Habs91} \citeyear{Habs91} and for SIS300 recently by
\citeauthor{Schramm04} \citeyear{Schramm04}, and
\citeauthor{Schramm06} \citeyear{Schramm06}.} Momentum can be
transferred to the Li-like ion by absorption of a laser photon as
well as by re-emission of the X-ray. While the former momentum
transfer is only $\hbar\omega_L/c$, the latter is according to
equation (\ref{energy-distribution}) in the order $\gamma^2
\hbar\omega_L/c$ since the angular distribution of the X-rays is
strongly peaked into forward direction, see subsection
\ref{SubSectAngDist}. The mean longitudinal momentum transfer in
the laboratory system due to absorption and emission of a photon
of energy $\hbar\omega_0$ in the rest frame of the Li-like system
is in the laboratory frame given by
\begin{eqnarray}
 \overline{\delta p_{||}} &=& \frac{\hbar\omega_L}{c} + \int\frac{(\hbar\omega_0/c)\cos\Theta}
 {\gamma(1-\beta\cos\Theta)}\cdot \frac{f(\cos\Theta_0)}{4\pi}
 \cdot d\Omega_0 = \nonumber\\
 &=& \frac{\hbar\omega_L}{c} + \int_{-1}^{1}\frac{(\hbar\omega_0/c)\cos\Theta}
 {\gamma(1-\beta\cos\Theta)}\cdot\frac{d(\cos\Theta)}{2\gamma^2(1-\beta\cos\Theta)^2}
 = \label{LongMomentum}\\
 &=& \frac{\hbar\omega_L}{c} + \beta\gamma\frac{\hbar\omega_0}{c} =
 \gamma\frac{\hbar\omega_0}{c}~.\label{LongMomentumResult}
\end{eqnarray}
Here $(\hbar\omega_0/c)\cos\Theta/(\gamma(1-\beta\cos\Theta)$ is
the longitudinal momentum transfer to the the Li-like ion in the
laboratory system by emission of a photon of momentum
$\hbar\omega_0/c$ in the rest frame of the ion. This relation
follows from equation (\ref{hbaromegaX}) for
$\overrightarrow{\psi} = 0$ and $\theta = \Theta$ after division
by $c$, which transforms the photon energy into the photon
momentum, and the projection of the momentum on the velocity axis
of the Li-like ion. The function $f(\cos\Theta_0) = 1$ represents
the angular distribution of the photon in the rest frame of the
Li-like ion which is assumed to be isotropic. In equation
(\ref{LongMomentum}) the solid angle is transformed by means of
equation (\ref{dOmegaTransform}) into the laboratory system in
which the integration is carried out. The integral of equation
(\ref{LongMomentum}) can be solved analytically. The final result
on the right hand side of equation (\ref{LongMomentumResult}) has
been obtained with equation (\ref{hbaromega0}) for $\Psi = 0$
after appropriate reshapings \footnote{This result can also
directly be obtained by a Lorentz transformation of the
transferred momentum $\hbar\omega_0/c$ in the rest frame of the
Li-like ion into the laboratory system.}. The instantaneous
cooling force is in the laboratory system at saturation $F_{||} =
dp_{||}/dt = \overline{\delta p_{||}}\cdot (1/2) \cdot
A_{ki}/\gamma = (1/2)\cdot A_{ki}\cdot\hbar\omega_0/c$.

The energy transfer to the Li-like ion of rest mass $M_0$ is
\footnote{Also this relation follows directly from the Lorentz
transformation formulas.} $\delta E_{||} = \delta(\gamma M_0 c^2)
= \beta\cdot\delta(\beta\gamma M_0 c)c =
\beta\cdot\overline{\delta p_{||}} =  \beta\gamma\hbar\omega_0/c$,
and the mean relative change of the relativistic factor at
resonance absorption and re-emission of a single photon is given
by
\begin{equation}\label{deltaGamma}
 \frac{\delta\gamma}{\gamma} =
 \beta\frac{\hbar\omega_0}{M_0 c^2}= (1+\beta)\beta\gamma\frac{\hbar\omega_L}{M_0 c^2}
 \cong 2\gamma\frac{\hbar\omega_L}{M_0 c^2}~.
\end{equation}
The numerical result for Li-like uranium at $\gamma = 25.68$ is
$\delta\gamma/\gamma = 1.27\cdot10^{-9}$. A total energy shift
$\Delta\gamma_{b}/\gamma = 10^{-4}$ requires a small de-tuning of
the laser frequency in such a manner that first only the ions with
largest energies are optically pumped. These are shifted to lower
energies by a gradual increase of the laser frequency. In this
manner successively more and more ions of the bunch are included
in the pumping process. For a complete cooling cycle at least a
number $(\Delta\gamma_{b}/\gamma)/(\delta\gamma/\gamma)$ of
transitions is required.

The rate at which $\gamma$ varies at saturation follows for a
pulse laser system with pulse width $\Delta t_{pulse}$ and
repetition rate $f_{rep}$ from equations (\ref{deltaGamma}) and
(\ref{Nsat}) as
\begin{equation}\label{rcool}
 \frac{\dot{\gamma}}{\gamma}\Big|_{sat} = \frac{d\gamma/\gamma}{dt}\Big|_{sat} =
 \beta\frac{\hbar\omega_0}{A\cdot m_u c^2}
 \frac{1}{2}\Big(A_{ki}\frac{\Delta t_{pulse}}{\gamma}+1\Big)\cdot f_{rep}~.
\end{equation}
Here $A$ is the atomic number of the Li-like ion and $m_u$ the
atomic mass unit. With the numbers of our example the cooling rate
is $\dot{\gamma}/\gamma |_{sat} = 0.46\cdot10^{-4}$/s for
$^{238}$U. A lower limit for the corresponding cooling time is for
$\Delta\gamma_{b}/\gamma = 10^{-4}$
\begin{equation}\label{tcool}
 \tau_c =
 \frac{\Delta\gamma_{b}/\gamma}{\dot{\gamma}/\gamma |_{sat}}
 = 2.2 \mbox{~s}~.
\end{equation}

It might be of interest to know how the cooling rate varies as
function of the charge number $Z$. For $^{120}$Sn ($Z$ = 50), as
an example, one obtains with $\hbar\omega_0 = 107.95$ eV and
$A_{ki} = 4.606\cdot 10^{9}$/s, both taken from
\citeauthor{Johnson96}, $\gamma = 25.68$, and $f_{rep}$ = 10 kHz a
$\dot{\gamma}/\gamma |_{sat} = 0.135\cdot10^{-4}$/s. However,
cooling of Li-like Sn via the $^{2}P_{3/2}-{}^{2}S_{1/2}$
transition at $\hbar\omega_0 = 374.70$ eV, $A_{ki} = 2.057\cdot
10^{11}$/s, and an increased $\gamma = 36$ results in a nearly two
orders of magnitude faster rate $\dot{\gamma}/\gamma |_{sat} =
9.74\cdot10^{-4}$/s.

While in longitudinal direction the ions are cooled, they are
heated in transverse direction. The variance of the transverse
momentum transfer is given by
\begin{equation}\label{delta-p-transverse}
 \overline{\delta p_{\bot}^2}=\int_{-1}^{1}\frac{(\hbar\omega_0/c)^2\sin^2\Theta}
 {\gamma^2(1-\beta\cos\Theta)^2}\cdot\frac{1}{2}\frac{d(\cos\Theta)}{\gamma^2(1-\beta\cos\Theta)^2}
 = \frac{2}{3}\Big(\frac{\hbar\omega_0}{c}\Big)^2~.
\end{equation}
Again, the integral has been solved analytically with the result
given at the right hand side of equation
(\ref{delta-p-transverse}) \footnote{The integration may as well
be performed in the rest frame of the Li-like system with the same
result $\overline{\delta
p_{\bot}^2}=\int_{-1}^{1}(\hbar\omega_0/c)^2\cdot\sin^2\Theta_0\cdot(1/2)\cdot
d(\cos\Theta_0)  = (2/3)(\hbar\omega_0/c)^2$.}. The angular spread
is
\begin{equation}\label{delta-p-transverse-rel}
 \frac{\sqrt{\overline{\delta p_{\bot}^2}}}{\beta\gamma M_0 c} =
\sqrt{\frac{2}{3}}\cdot\frac{\hbar\omega_0}{\beta\gamma M_0 c^2}
\end{equation}
with $\beta\gamma M_0 c$ the momentum of the Li-like system. The
total angular spread after a number of
$(\Delta\gamma_{b}/\gamma)/(\delta\gamma/\gamma)$ uncorrelated
emissions of photons is
\begin{equation}\label{angular-spread}
 \sigma^{\prime}_{HI} = \sqrt{\frac{\Delta\gamma_{b}/\gamma}{\delta\gamma/\gamma}}
 \sqrt{\frac{2}{3}}\cdot\frac{\hbar\omega_0}{\beta\gamma M_0 c^2}~.
\end{equation}
With the numbers of our example the result is
$\sigma^{\prime}_{HI} = 0.0113~\mu$rad which probably is
negligibly small in comparison to the angular spread of the HI
beam with an emittance of about 1 $\pi$ mm mrad.

Since the energy spread after laser cooling is in the order
$\Delta\nu_L/\nu_L = 5.3\cdot 10^{-7}$, a superior line width in
the hyperfine structure pattern can be expected with laser
cooling.
\section{Nuclear polarization by optical pumping} \label{pumping}
\begin{figure}[t,b]
\centering
\includegraphics[scale=0.5,clip]{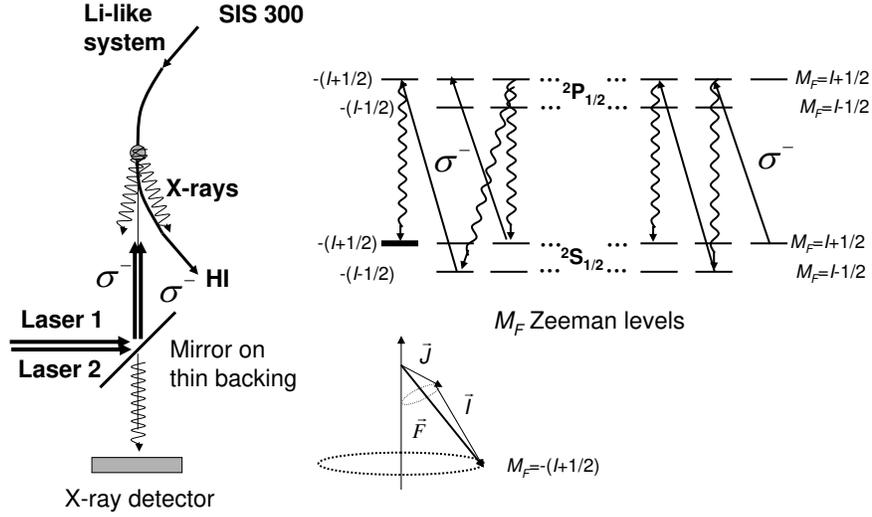}
\caption{Scheme of the experimental setup for polarization of the
Li-like ion beam (left), and pumping scheme with circularly
polarized laser light (right). Shown are the Zeeman levels for the
hyperfine levels of the ground and excited state with $F =
I\pm1/2$, with $I$ the nuclear spin. After a very long pumping
time only the $F = -(I+1/2)$ Zeeman level will be populated.}
\label{SetupOptPump}
\end{figure}

A few remarks will be added on the possibility to prepare a
polarized Li-like ion beam which exhibits also a nuclear
polarization if the nuclear spin is not zero. As schematically
shown in figure \ref{SetupOptPump}, pumping with left-circularly
polarized laser light results after many transitions finally in a
population of only the $|F=I+1/2,M_F = -(I+1/2)\rangle$ Zeeman
level, because this level can not be pumped anymore. In this state
the electronic spin as well as the nuclear spin are polarized as
sketched schematically in the vector coupling model inset in
figure \ref{SetupOptPump}.

Unfortunately, a polarization may probably not be maintained
within the SIS300 ring since in the strong magnetic field of the
bending magnets and beam optical elements a polarized total
angular momentum $\overrightarrow{F}$ precesses and may randomize.
However, it might be conceivable that an external Li-like beam may
be pumped in a long straight section, for instance by 100 ns long
circularly polarized laser beams which would induce, according to
equation (\ref{Nsat}), $N_{0,sat} = 32$ transitions in an
unpolarized Li-like ion. This is a rather large number and the
attained polarization may already be high. For Li-like ions with
$Z\leq 50$, it could be pumped via the much faster
$^{2}P_{3/2}-{}^{2}S_{1/2}$ transition and even much larger
numbers may be attained at even shorter laser pulse durations. For
quantitative numbers, however, detailed calculations are required
which were beyond the scope of this work.
\section{Conclusions} \label{conclusions}
Precision transition energies $\hbar\omega_0$ as well as
relativistic factors $\gamma$ can be measured for Li-like ions at
the future heavy ion accelerator SIS300 if the inherently very
precise laser spectroscopy method is combined with a precision
X-ray energy measurement of fluorescence photons by means of a
single crystal monochromator. Isotope shifts and magnetic moments
can be measured at the $^{2}P_{1/2}-{}^{2}S_{1/2}$ transition for
all elements, and quadrupole moments for Z$\leq$50 at the
$^{2}P_{3/2}-{}^{2}S_{1/2}$ transition, in principle even at a
single stored radioactive species far off stability if their
nuclear spin is known. Heavy ion beams in SIS300 may be laser
cooled, and polarized external heavy ion beams may be prepared by
optical pumping with polarized laser beams.


\begin{acknowledgements} I thank Dr. W. Lauth, Dr. A. Wolf for
fruitful discussions, Dr. P. Kunz for critical comments on the
manuscript, and Dr. F. Hagenbuck for valuable information on
SIS300.

This work has been supported by Bundesministerium f\"{u}r Bildung
und Forschung under contract 06 MZ 169 I. \end{acknowledgements}

\end{article}

\end{document}